# Spectroscopic Detection of a Cusp of Late-type Stars around the Central Black Hole in the Milky Way

M. Habibi[1], S. Gillessen[1], O. Pfuhl[1], F. Eisenhauer[1], P. M. Plewa[1], S. von Fellenberg[1], F. Widmann[1], T. Ott[1], F. Gao[1], I. Waisberg[1], M. Bauböck[1], A. Jimenez-Rosales[1], J. Dexter[1], P. T. de Zeeuw[1,2], and R. Genzel[1,3]
[1] Max-Planck-Institut für Extraterrestrische Physik, D-85748 Garching, Germany
[2] Sterrewacht Leiden, Leiden University, Postbus 9513, 2300 RA Leiden, The Netherlands
[3] Physics Department, University of California, Berkeley, CA 94720, USA

## Abstract

In a dynamically relaxed cluster around a massive black hole a dense stellar cusp of old stars is expected to form. Previous observations showed a relative paucity of red giant stars within the central 0.5 pc in the Galactic Center. By co-adding spectroscopic observations taken over a decade, we identify new late-type stars, including the first five warm giants (G2-G8III), within the central 1 arcsec$^2$ (0.04 × 0.04 pc$^2$) of the Galaxy. Our findings increase the number of late-type stars to 21, of which we present deep spectra for 16. The updated star count, based on individual spectral classification, is used to reconstruct the surface density profile of giant stars. Our study, for the first time, finds a cusp in the surface number density of the spectroscopically identified old (>3 Gyr) giants population ($m_K < 17$) within 0.02–0.4 pc described by a single power law with an exponent $\Gamma = 0.34 \pm 0.04$.

## 1. Introduction

Located at a distance of only 8 kpc and containing a supermassive black hole (SMBH) of mass $4.1 \times 10^6 M_\odot$ (Boehle et al. 2016; Gillessen et al. 2017; Gravity Collaboration et al. 2018), the nuclear star cluster (NSC) of the Milky Way is of exceptional interest for exploring the impact of an SMBH on its surrounding stellar population.

Observations of the central few parsecs of the Milky Way have revealed a complex stellar population that is mostly old, but has experienced several star formation events in perhaps a continuous but episodic limit cycle (Morris & Serabyn 1996; Blum et al. 2003; Genzel et al. 2010; Pfuhl et al. 2011; Schödel et al. 2018). Young stars in the NSC show an increasing density toward SgrA* (Paumard et al. 2006; Do et al. 2009; Lu et al. 2009, 2013; Bartko et al. 2010). The origin of the young stars, especially those in the immediate vicinity of the SMBH, is puzzling given the strong tidal shear that prevents the gravitational collapse (Morris 1993). Observing a population of giants is expected in an old NSC. However, number counts of the brightest giant stars are inconsistent with the predicted properties of an equilibrium "stellar cusp" around a black hole (BH; e.g., Buchholz et al. 2009; Do et al. 2009; Bartko et al. 2010). Recently, in a series of three papers the density profile of late-type stars is revisited via a photometric number count study and a diffuse light analysis (Gallego-Cano et al. 2018; Schödel et al. 2018; Baumgardt et al. 2018). Their studies find a cuspy profile for faint giants, and possibly sub-giants and main sequence stars. Consistent with previous studies, they also reported a core-like surface density of red clump and brighter giant stars ($m_K < 16$) within the central 0.3 pc.

Different scenarios are proposed to explain the apparent deficit of brightest giants: tidal disruption of giants by the SMBH (Hills 1975), collisional destruction with a clumpy gas disk (Amaro-Seoane & Chen 2014), and collisional destruction with other stars and remnants (Alexander 2005; Davies & King 2005; Dale et al. 2009). Some of the proposed scenarios to remove/alter the population of giants were also motivated to explain the surprising presence of young stars in the region. For example, it was suggested that young S-stars[4] are the stripped cores of luminous asymptotic giant branch stars. Analyzing K-band (Eisenhauer et al. 2005; Martins et al. 2008), and later combining 12 yr of K- and H-band observations of the young S-stars, shows that they are high surface gravity (dwarf) stars, excluding the stripped-giant scenario for the young S-stars (Habibi et al. 2017).

Due to the high extinction and extreme stellar crowding, it is observationally challenging to confirm whether or not a stellar cusp exists. The crowding in the central regions, as well as the increasing density of young stars toward the center, can confound the measured density of late-type stars around the SgrA* (Genzel et al. 2003; Schödel et al. 2007). The intrinsic H–K color of the stars in the inner 1″ is smaller than 0.4 mag, which makes it hard to distinguish the un-relaxed young stellar population from the possible old cusp structure. The stellar distribution needs to be corrected for the spectroscopically identified young stars. All existing studies of the distribution of giants around SgrA* had to rely on the available relatively shallow spectroscopy, even toward the central few arcseconds, in order to exclude younger stars from their analyses.

Adaptive optics (AO) spectroscopy allows one to observe stars down to 15.5–16 $m_K$ (Do et al. 2009; Bartko et al. 2010; Pfuhl et al. 2011; Habibi et al. 2017). In this Letter, we analyze the rich Galactic Center (GC) data set, observed after the latest spectroscopic study of the distribution of giants by Bartko et al. (2010), and expand the approach of combining multi-epoch AO-assisted spectroscopic observations of the GC (Martins et al. 2008; Habibi et al. 2017) to study the late-type stars in the vicinity of SgrA* down to 17.5 mag. High signal-to-noise ratio (S/N; S/N > 20–120) spectra enable us to assign temperatures and ages to the closest giants around the SMBH including the fainter late-type stars. All cusp formation models assume that the stellar cluster is dynamically relaxed. The unique advantage of our spectroscopic study is that through an individual age estimation, we can select a stellar population that is old enough (older than the GC dynamical relaxation timescale ∼3 Gyr, Alexander 2017) to have undergone dynamical relaxation. We can also study the unexplored range of 15–17 $m_K$ within the central arcsecond. This magnitude range is critical both for exploring the stellar cusp and the star formation history of the GC.

---

[4] The term "S-star" has been used to refer to stars around the SMBH, mostly within 1 arcsec radius from SgrA*, for which one can hope to determine orbits. These stars probably belong to a distinct dynamical population with randomly aligned orbits (Gillessen et al. 2017), and as this Letter emphasizes they are not necessarily young stars.

This Letter is organized as follows. Section 2 lists the observations and briefly describes the reduction of the data to obtain deep spectra. In Section 3, we present the spectral classification of late-type S-stars as well as their evolutionary stage on a Hertzsprung–Russell (H–R) diagram based on spectroscopic and photometric observations. In this section, we also report on the discovery of the first G-type stars within the central arcsecond. Based on the estimated stellar parameters in Section 3, we calculate the tidal disruption radius of stars in Section 4. Section 5 presents an updated projected distribution of late-type stars within the central 0.5 pc and discusses the theoretically expected cusp around the SMBH.

## 2. Observation and Data Reduction

In this Letter we carried out an analysis on AO-assisted spectroscopic data observed with SINFONI during over a decade (2006–2018) and partly published in previous studies (e.g., Eisenhauer et al. 2003, 2005; Gillessen et al. 2017; Habibi et al. 2017). All epochs are observed using the $H+K$-band grating with spectral resolution of ∼1500, and the exposure time of 600 s per frame. A summary of the data used in this study and the epochs used to extract each star's spectrum are given in Table 1. We refer the reader to Habibi et al. (2017) for more details on the reduction steps.

We chose epochs in which the star of interest is unconfused with other bright stars. We construct three deep SINFONI master cubes based on observations during years 2013, 2014, and 2018. The extracted spectrum from these master cubes are used as the first guess for the spectral type investigation. We confirm the first guess through combining individual spectra from different epochs and constructing deep spectra. To normalize the spectra we use the minimum component filter technique (Wall 1997). We cross-correlate the CO bandheads with a template spectrum to measure the radial velocities and Doppler-shift the spectra to rest wavelength. Multiplicity of the CO band features helps us to eliminate misidentification in noisy spectra. If a prominent Brγ line is detected, we use both the Brγ and CO bandheads to measure the velocity. In an iterative process the measured velocities are refined by cross-correlating each individual spectrum with the preliminary combined spectrum used as a template (see Section 2.4 of Habibi et al. 2017). A final deep spectrum is made with an S/N weighted average of all the individual spectra at the rest wavelength. We cross-check the final deep spectrum with the extracted spectrum from our master epochs to eliminate a possible misidentification. The consistency of estimated radial velocities from different epochs guarantee that we do not cross-correlate random noise patterns in different epochs. Prominent Mg I, Ca I, Na I, and Fe lines appear in the combined spectrum of all the analyzed late-type S-stars, which justifies the method that we use.

**Table 1**
Summary of SINFONI Observations of the Central Arcsecond used to Construct the Deep Spectra in this Study

| Date | Band | $t_{\rm exp}$ on S2 [min] | FWHM [mas] | Stars |
|---|---|---|---|---|
| 2006 Mar 17 | $H+K$ | 8.3 × 10 | 70 | 24, 27, 25 |
| 2007 Jul 18–23 | $H+K$ | 10 × 14 | 70 | 27 |
| 2008 Apr 6 | $H+K$ | 10 × 21 | 56 | 20, 35, 21, 27, 38, 43 |
| 2009 May 21–25 | $H+K$ | 10 × 11 | 64 | 20, 37 |
| 2010 May 10–12 | $H+K$ | 10 × 26 | 58 | 20, 37, 38, 43 |
| 2011 Apr 11–27 | $H+K$ | 10 × 30 | 55 | 134 |
| 2011 May 2–14 | $H+K$ | 10 × 11 | 53 | 134 |
| 2012 Jun 26–Jul 01 | $H+K$ | 10 × 28 | 58 | 10, 35, 17, 134 |
| 2012 Jul 07–11 | $H+K$ | 10 × 32 | 64 | 10, 35, 17, 21 |
| 2013 Apr 5–18 | $H+K$ | 10 × 121 | 60 | 20, 10, 35, 41, 43, 17, 21, 51, 27, 32, 38, 37, 134, 36 |
| 2013 Aug 28–31 | $H+K$ | 10 × 26 | 57 | 20, 35, 134 |
| 2013 Sep 18–26 | $H+K$ | 10 × 19 | 74 | 20 |
| 2014 Apr 5–8 | $H+K$ | 10 × 18 | 59 | 20, 32, 38, 134 |
| 2014 Mar 28–Apr 10 | $H+K$ | 10 × 33 | 37, 56 | 20, 32, 38 |
| 2014 Apr 22–24 | $H+K$ | 10 × 27 | 54 | 20, 32, 37, 38, 134 |
| 2014 May 8–9 | $H+K$ | 10 × 29 | 54 | 20, 41 |
| 2014 May 27–Jun 10 | $H+K$ | 10 × 6 | 102 | 20, 41 |
| 2014 Jul 08–20 | $H+K$ | 10 × 33 | 67 | 20, 51, 41 |
| 2014 Aug 18–31 | $H+K$ | 10 × 32 | 70 | 20, 51, 41 |
| 2015 Apr 20–29 | $H+K$ | 10 × 43 | 63 | 20, 10, 17, 21, 51 |
| 2015 May 18–23 | $H+K$ | 10 × 9 | 71 | 20, 51 |
| 2016 Apr 14–16 | $H+K$ | 10 × 19 | 56 | 20, 10, 35, 17, 21, 25, 32, 37, 134 |
| 2016 Jul 9–11 | $H+K$ | 10 × 3 | 64 | |
| 2018 4–5 | $H+K$ | 10 × 82 | 51 | 43, 30 |

**Note.** The stars for which we have used a particular epoch are given in the fifth column. The on-target exposure times refer to S2.

## 3. From Deep Spectra to H–R Diagram

Figure 1 shows the combined spectra with an S/N of 20–120 in the $K$-band. The most prominent spectral features are the $^{12}$CO (2,0) bandhead (2.29–2.5 μm), Na I doublet (2.2062 and 2.2090 μm), Ca I doublet (2.263 and 2.267 μm), and Mg I (2.280 μm). These lines are complex blends with Sc, Ti, V, and Fe, which is in agreement with previous observations of late-type stars (e.g., Wallace & Hinkle 1997; Rayner et al. 2009). These spectral features are typical of G-M stars. We also observe a prominent Brγ line in one to four of the stars.

The CO bandheads are good indicators of the stellar effective temperature ($T_{\rm eff}$, Kleinmann & Hall 1986; Frogel et al. 2001; Pfuhl et al. 2011). Schultheis et al. (2016), who explored the giants within the temperature range 3200 K < $T_{\rm eff}$ < 4500 K, did not find any dependence of the $T_{\rm eff}$–EW$_{\rm CO}$ relation on the metallicity of the star, which makes it an attractive tool for Galactic bulge studies. We measured the equivalent width of CO bandheads (EW$_{\rm CO}$) according to the recipe of Frogel et al. (2001). To measure the effective temperature we use the $T_{\rm eff}$–EW$_{\rm CO}$ relation from Pfuhl et al. (2011).

To determine bolometric magnitudes, we derive the absolute magnitudes by adopting the GC distance of 8.1 kpc

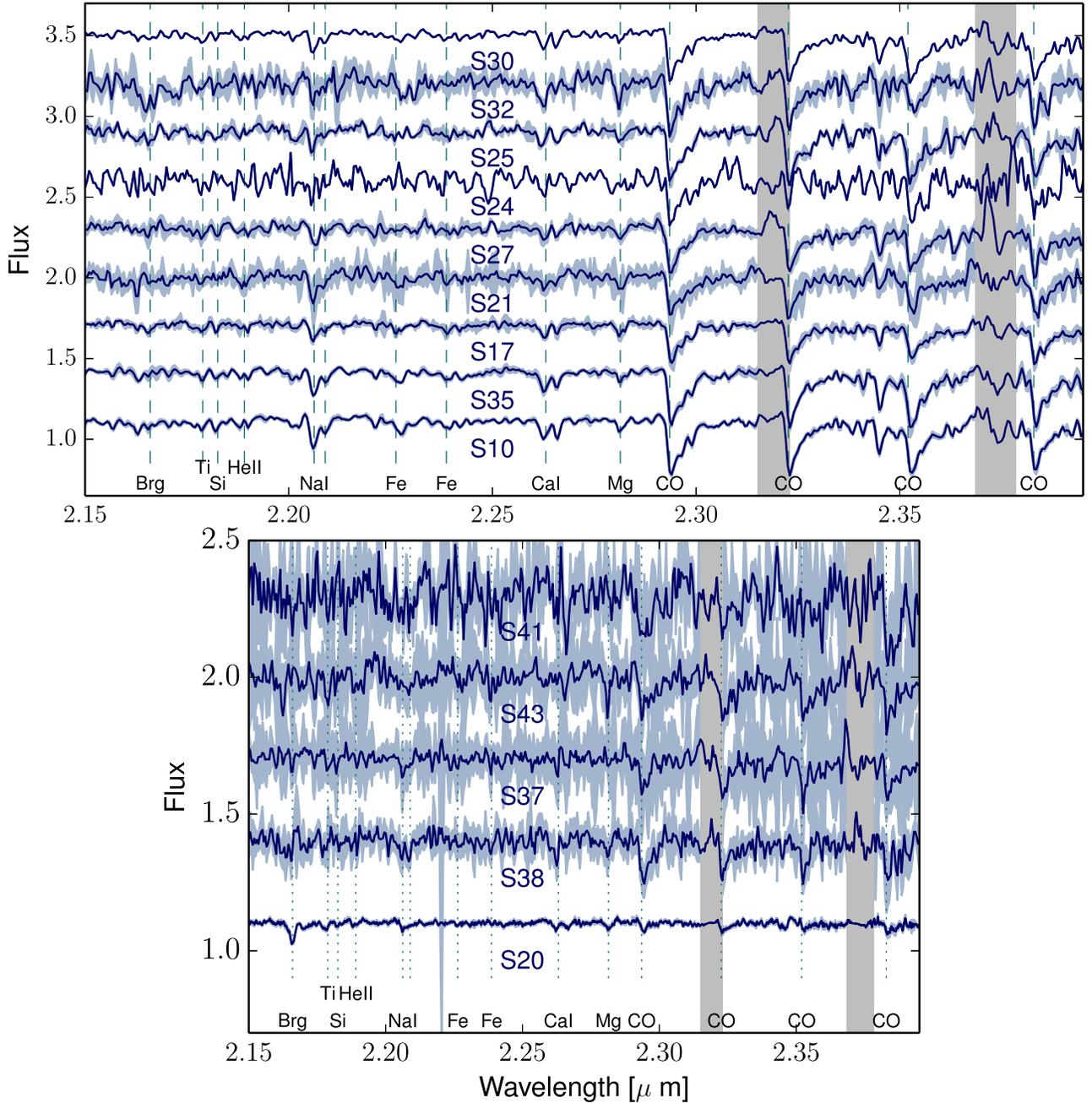

**Figure 1.** Top panel: combined spectra of eight K0-K3III-type stars ($T_{\rm eff} \sim$ 4100–4600 K) within the central 0.2 pc (the spectra of S134 and S51 are not shown on this scale due to higher noise level). Bottom panel: combined spectra of the first five warm giants of G2-G8III type ($T_{\rm eff} \sim$ 4700–5550 K) within 1″. The shaded region around the spectrum illustrates the standard error of the mean calculated for combined epochs. Vertical gray shaded regions mark areas in which we have strong contamination from uncorrected sky emission (2.315–2.323 and 2.368–2.375 μm).

(Gravity Collaboration et al. 2018) and the $K$-band extinction in the central arcseconds determined by Fritz et al. (2011), $A_K = 2.4$. The effective temperature is applied to obtain the $K$-band bolometric correction, $BC_K$, using the relation

$$BC_K = 2.6 - (T_{\rm eff} - 3800)/1500 \quad (1)$$

from Blum et al. (2003). The resulting location of the stars on the H–R diagram is shown in Figure 2. The error bars are calculated considering both the formal fit uncertainties of applied $T_{\rm eff}$–$EW_{\rm CO}$ relation, and also the inherent rms scatter of such a calibration fit ~100 K (Pfuhl et al. 2011; Schultheis et al. 2016).

We compared $EW_{\rm CO}$ of our stars with the spectral-type–$EW_{\rm CO}$ calibration relationship derived by Negueruela et al. (2010), using the G0-M7 stars (atlas of infrared spectra; Rayner et al. 2009). This reveals the spectral-type of M1III for S134 and K0-K3 III for 10 analyzed stars: S10, S17, S21, S51, S134, S24, S25, S27, S30, S32, and S35 ($T_{\rm eff} \sim$ 4100–4600 K). We determine a G2-G8III spectral-type for the stars: S20, S38, S37, S41, and S43 ($T_{\rm eff} \sim$ 4700–5550 K). The spectral-type for the S41, S43, S51, S20, and S134 stars are assigned for the first time (see Table 2). The stars S37 and S20 (two of the warmest giants in the region) were initially thought to be early-type stars, which shows the necessity of high S/N spectra to detect the weak late-type features of these stars. Photometric identification of these stars ($m_K$:

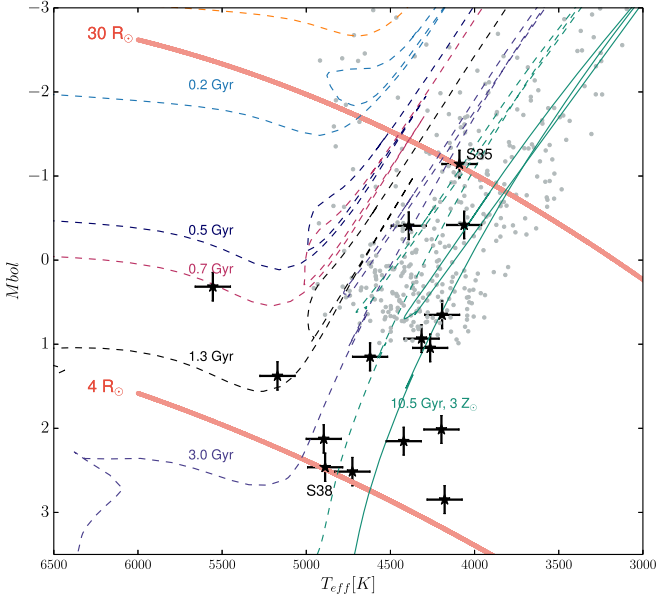

**Figure 2.** H–R diagram of late-type stars within the central 0.02 pc (black point with error bars). Over-plotted are theoretical isochrones ($Z = 1.5\,Z_\odot$) for 0.2–10 Gyr. Also shown is a super-solar isochrone ($Z = 3\,Z_\odot$, the solid line) to explain the location of cool and fainter stars on the diagram. The gray dots show the cool GC stars from Pfuhl et al. (2011). Two pink arcs show curves of constant radii for 4 and 30 $R_\odot$, which are the least and most extended atmospheres we observe in this region.

**Table 2**
List of Spectrally Identified Giant Stars within the Central Arcsecond

| Star ID | $\Delta x$ [″] | $\Delta y$ [″] | $m_K$ | Old SP | New SP |
|---|---|---|---|---|---|
| 30 | 0.538356 | 0.396562 | 14.3 | Late-type | K3III [†,∗] |
| 35 | −0.567608 | −0.428963 | 13.4 | Late-type | K5III [†,∗] |
| 21 | 0.348759 | −0.125537 | 16.9 | Late-type | K2III [†,∗] |
| 20 | −0.218038 | 0.073692 | 15.8 | Early-type | G2III [∗] |
| 43 | 0.461173 | −0.119937 | 17.5 | Unknown | G7.5III [∗] |
| 51 | 0.434546 | −0.288149 | 17.4 | Unknown | K3.5III [∗] |
| 37 | −0.325299 | 0.407447 | 16.6 | Early-type | G5.5III [†,∗] |
| 27 | −0.164155 | 0.529491 | 15.6 | Late-type | K3III [†,∗] |
| 134 | −0.279500 | 0.448500 | 17.3 | Unknown | M1III [∗] |
| 36 | −0.286713 | 0.220915 | 16.4 | Unknown | Late-type [†,∗] |
| 38 | 0.214675 | 0.013131 | 17 | Late-type | G7.5III [†,∗] |
| 50 | 0.498113 | −0.509483 | 17.2 | Unknown | Late-type [†] |
| 32 | 0.325010 | −0.371383 | 16.6 | Late-type | K3.5III [†,∗] |
| 41 | 0.201996 | −0.325967 | 17.4 | Unknown | G9III [∗] |
| 25 | 0.104875 | −0.435053 | 15.2 | Late-type | K3.5III [†,∗] |
| 24 | 0.114017 | −0.486130 | 15.7 | Late-type | K3III [†,∗] |
| 10 | −0.037957 | −0.376817 | 14.1 | Late-type | K5III [†,∗] |
| 45 | −0.184699 | −0.549444 | 15.6 | Late-type | Late-type [†] |
| 46 | −0.261075 | −0.552917 | 15.6 | Unknown | Late-type [†] |
| 34 | −0.380040 | −0.461133 | 15.5 | Late-type | Late-type [†] |
| 17 | −0.062286 | −0.004979 | 16.0 | Unknown | K0III [†,∗] |

**Note.** The first three columns present the stellar ID and position of the stars in arcsecond offsets ($x$ is aligned with R.A. and $y$ with decl.) from the SgrA$^*$ for the epoch 2009.0 in accordance with Figure 3. The $K$-band magnitudes are given in the fourth column. The previous spectral classification (SP; Gillessen et al. 2009), which is used for the previous study on the stellar density profile (Bartko et al. 2010), is given in the fifth column. The new spectral classification based on Gillessen et al. (2017)[†] and this study[∗] are shown in the last column. The late-type stars are the giants for which the CO bandhead are observed, but pinpointing the exact spectral type was not possible due to a noisy spectrum. The early-type stars are usually of type BV (Eisenhauer et al. 2005; Habibi et al. 2017).

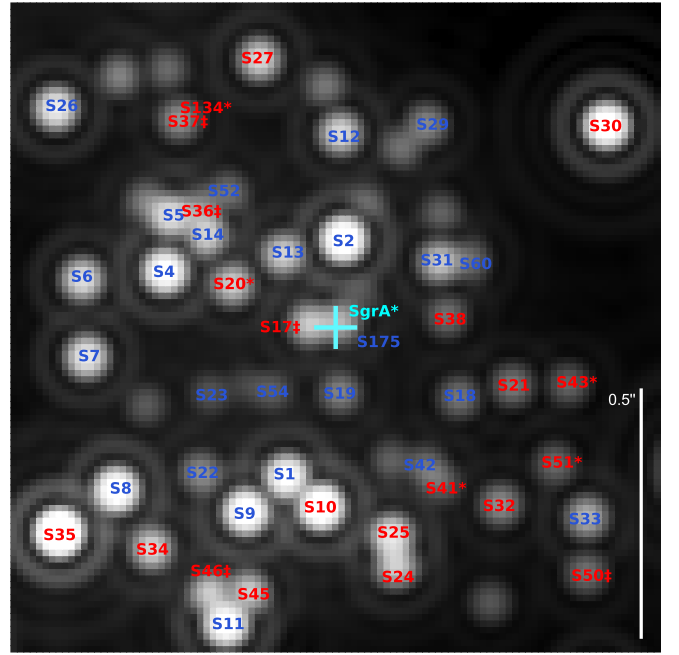

**Figure 3.** Finding chart of the spectroscopically identified S-stars within the central arcsecond for the epoch 2009.0. The mock image is constructed based on our NACO data. The blue stars are the younger dwarf stars, and the red stars are the giants. The red stars marked with ‡ and ∗ are the newly identified late-type stars in Gillessen et al. (2017) and this study, respectively. The rest of giants are the previously known late-type S-stars.

15–17) requires a photometric accuracy that is better than 5%. Previous photometric studies such as Buchholz et al. (2009) could be partly affected by this limit. Perhaps due to similar overlooked classification limitations, all of the early-type candidates suggested by Nishiyama & Schödel (2013) were later identified as giants of intermediate temperature (Nishiyama et al. 2016). The finding chart of young and late-type S-stars within the central arcsecond is shown in Figure 3.

Most of the late-type S-stars are consistent with 3–10 Gyr slightly super-solar isochrones (Figure 2). However, few of the coolest observed S-stars can only be explained by super-solar metallicity isochrones. Previous studies have reported observing metal-rich stars in the GC (Cunha et al. 2007; Do et al. 2015; Feldmeier-Krause et al. 2017; Rich et al. 2017). Pfuhl et al. (2011) and Nishiyama et al. (2016) observed similar stars within the NSC and described them as possible outliers due to different extinction or high metallicity. We do not expect strong extinction variations within the 0.04 pc (1″) scale. The apparent spread of late-type S-stars on the H–R diagram probably implies a wide metallicity range for these stars. We are building up the S/N of these stars to explore their metallicity spread in a future study.

Comparing their position on the H–R diagram with the theoretical isochrones (Chen et al. 2014; Tang et al. 2014), we estimate the mass and radius range for these stars. Assuming a GC metallicity of $Z = 1.5\,Z_\odot$ (Cunha et al. 2007) the estimated initial-masses lie within 0.5–2 $M_\odot$ and radii of these stars are within 4–30 $R_\odot$. Assuming a different metallicity will move the theoretical isochrone mostly along the horizontal axis and will not affect the radius measurements (see the schematic constant-radius curves in Figure 2). The individual stellar parameters for these stars can only be meaningfully assigned after spectral modeling of their deep combined spectra and comparing all available observables simultaneously to stellar models through a Bayesian statistical method.

## 4. Tidal Disruption Radius of Late-type Stars

Based on the estimated stellar parameters (Section 3), we calculate a tidal disruption radius, $r_t$, range of $0''.0004$–$0''.003$ for the late-type S-stars ($mk < 17.5$) in the central projected arcsecond. A star is tidally disrupted if $r_t$ is larger than the Schwarzschild radius and its orbital angular momentum is small enough so that orbital periapse is less than the tidal radius (Alexander 2005). With the SMBH Schwarzschild radius of $\sim 10$ microarcsec, the first condition holds for the calculated $r_t$ of all the stars. The second criterion for a tidal disruption event is not fulfilled for the observed giant stars. The $r_t$ of the shortest period late-type star, S38, is $0''.0004$, which is 100 times smaller than the star's periapse distance (Gillessen et al. 2017). Although we have not determined an orbit for most of these late-type stars, we can compare the largest measured $r_t$ ($0''.003$ for S35) with the periapse distance of the closest well-measured orbit, i.e., S2. The largest $r_t$ in this region is five times smaller than the periapse distance of the star S2 ($\sim 0''.015$). All of the existing (remaining) giants in the vicinity of the SMBH ($mk < 17.5$) are stable against tidal disruption by the SMBH, considering the current orbits of early- and late-type S-stars.

## 5. Surface Density of the Inner Cusp

The newly identified late-type stars in Gillessen et al. (2017) and the stars found in this study add about 10 stars to the previously known sample of late-type stars within the central arcsecond (Figure 3). Our star count with a detailed spectral classification (see Section 3) within the central arcsecond$^2$ is not only photometrically complete at 17 mag, we also can identify the spectral type of all the stars with $m_K < 17$, irrespective of their stellar type. Although imaging within the crowded central arcsecond is confusion limited, monitoring the field for few decades and having high-precision imaging from NACO as a guide allows to partly overcome the confusion limit. We added artificial stars with proper motion typical for a circular motion around SgrA$^*$ to the spectrally integrated SINFONI images, and determined the probability of re-detecting the artificial stars in one or more epochs. We can re-detect 90% of the artificial stars.

Bartko et al. (2010) presented the projected-, completeness-, and coverage-corrected surface density profiles for all stars (without spectral classification) with $m_K < 17$ out to 10 arcsec from the center. The fraction of young stars is below 1 star/arcsec$^2$ within 2–10 arcsec. We correct the distribution of stars with $m_K < 17$ for the fraction of young stars, and build a distribution of late-type stars out to larger radii around the SgrA$^*$.

The confusion-corrected stellar number density within the central arcsecond along with reanalyzed density profile of stars with $m_K < 17$ from Bartko et al. (2010) (corrected for the fraction of young stars) is used to reconstruct the radial surface density of the giant stars $m_K < 17$ (Figure 4). We fit a power-law function, $\Sigma(R) \propto R^{-\Gamma}$, to this surface number density profile, where $\Sigma$ is the surface number density, $R$ the projected radius, and $\Gamma$ the power-law index. The resulting density profile has a power-law slope of $0.34 \pm 0.04$. The uncertainty of fit parameters is estimated from the covariance matrix of the fit using a jack-knife re-sampling method. The measured slope excludes a core-like distribution (blue dashed line in Figure 4) for stars with $m_K < 17$, within 0.02–0.4 pc, with $6.5\sigma$ significance. Our fitted value for the 2D power-law index perfectly matches with the measured surface density by Gallego-Cano et al. (2018), who found $\Gamma = 0.34 \pm 0.03$ for star counts at $16.5 < K_s < 17.5$ outside the central arcsecond ($0.05 < r < 0.4$ pc; see their Figure 13).

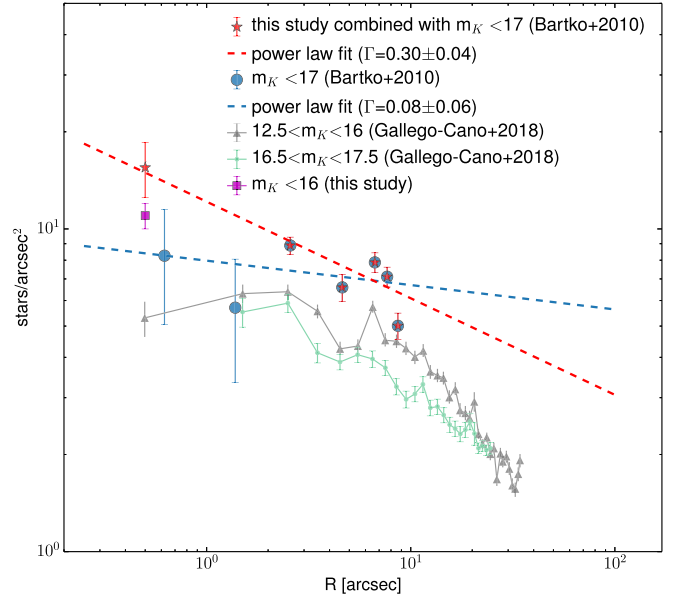

**Figure 4.** Surface density profiles of late-type stars ($m_K < 17$) out to distances of 10 arcsec. For the central 1 arcsec (red star), we use our star count with 100% photometric and spectral classification completeness. For larger radii (2–10 arcsec) around SgrA$^*$ we use the data from Bartko et al. (2010) corrected for the small fraction of young stars. The radial bin size is 1 arcsec. The dashed lines show power-law fits to the late-type stars ($m_K < 17$), including our updated star count (red) and without that (blue), respectively. The red power-law fit with the index of $\Gamma = 0.34 \pm 0.04$ reveals a cusp distribution in the region. Our updated star count for stars with $m_K < 16$ is marked with a magenta square. The gray and green points illustrate distribution of the bright and faint giants from Gallego-Cano et al. (2018).

To detect a possible dynamically relaxed stellar cusp around SgrA$^*$, one needs to probe stars that are older than the dynamical relaxation timescale ($\sim 3$ Gyr, Alexander 2005). Previous studies were limited to individual spectroscopic stellar classification for stars brighter than 15.5 $m_K$. The critical advantage of our observation within the central 0.04 pc is that we can estimate a stellar age for each star (see Figure 2). Therefore, we can verify that our observed sample solely consists of several Gyr old stars. Figure 2 shows that the two warmest giant stars in our sample are probably younger than 2 Gyr. Our preliminary spectral modeling analysis shows that indeed these stars are of low metallicity and do not belong to an intermediate age population. Nonetheless, we did the exercise of excluding the two stars from the surface density distribution and repeated the fit. The resulting power-law fit has an index of $\Gamma = 0.30 \pm 0.04$, which still excludes a core-like structures with $5.5\sigma$.

For a power-law cluster with infinite extent, the 3D power law index ($\gamma$ in $\rho \propto r^{-\gamma}$) equals to $\Gamma + 1$. This does not exactly hold for the NSC, due to the finite structure of the cluster. Schödel et al. (2018) and Gallego-Cano et al. (2018) fit 3D Nuker models to data with a similar $\Sigma$ values to what we measure in this Letter and yield a 3D power-law index of $\gamma = 1.1$–$1.4$. This is a clear cusp structure, which is somewhat flatter than the theoretically expected value of $\gamma = 1.75$–$2.3$ for a single-mass dynamically relaxed stellar population around a SMBH (e.g., Bahcall & Wolf 1976). Theoretical studies have progressed to consider the complex stellar population properties such as stellar mass distribution and mass segregation. They predict a mass-dependent cusp near the SMBH (Murphy et al. 1991; Alexander & Hopman 2009; Baumgardt et al. 2018). Within a multiple mass population, lighter stars can have a cusp slope as shallow as $\gamma = 1.5$ (Alexander & Hopman 2009). The most recent N-body simulation of star clusters around SMBHs consider the complex star formation history of the NSC and find a stellar cusp with 2D power-law index of $\Gamma \approx 0.3$ (Baumgardt et al. 2018).

Including the fainter stars, we do not observe a general lack of giants at projected distances of 0.02–0.4 pc from SgrA*. However, the problem of missing brightest giants still exists. Our study can only present an updated number count for the innermost data point of the stellar 2D distribution. The issue of missing brightest giants is very sensitive to this innermost data point and to the magnitude range that is used to define the brightest bin.

We identify four bright giants ($m_K < 15.5$) within the projected central arcsecond (see Table 2), which is two times larger than the previously measured value for the same magnitude range (Bartko et al. 2010). Our updated number count is still consistent with the existence of a core structure toward the center, but not with a central hole. Assuming that brighter giants follow the same distribution as fainter ones ($m_K < 17$), we estimate the number of missing giants ($m_K < 15.5$) within the central arcsecond$^2$ to be $\sim 4$ stars. As shown by Alexander & Hopman (2009), we expect to observe a mass-dependent cusp near an SMBH. Assuming up to 0.2–0.3 dex flatter 3D profile for giants of few solar masses, i.e., $\rho \propto r^{-1.5}$, the number of missing giants in this region increases to $\sim 5$ stars.

If we define the brightest magnitude bin according to the study by Do et al. (2009), i.e., $m_K < 16$, we find 11 star/arcsec$^2$, which is 1.5–2 times larger than their measured value. The updated number count is marked in Figure 4. Including this data point, one can fit a power law with a non-zero index to the existing stellar surface density and conclude that we start to see a weak cusp already at this magnitude range. Nonetheless, the 2D stellar density in this brightness range has a more complex structure than a simple power-law distribution.

The cusp slope of the cluster core can become shallower than $\gamma = 1/2$ if the physical collisions are frequent (Murphy et al. 1991; Alexander 1999). Despite various theoretical studies, up to now the effectiveness of stellar collisions in rearranging the stellar population is mostly uncertain (see Genzel et al. 2010). Similar to stellar collisions, extreme tidal interactions with the SMBH can, in principle, lead to the selective destruction of giant stars (Alexander 2003; Ghez et al. 2009). A study by Murphy (2011) used Fokker–Planck simulations and showed that for massive nuclei (10–100 times more massive than GC) stellar collisions play a much more important role; however, the GC model shows a flattening in the projected giants density slope due to the extreme tidal interactions with the SMBH. Although the conditions in which tidal interactions can happen have been discussed (Alexander & Morris 2003; Alexander 2005), their role in the resulting observed cusp has not been explored extensively.

Including the fainter stars, we find a cusp structure for the old giant stars ($m_K < 17$) within 0.25–10 arcsec (0.02–0.4 pc). Our observations reveal that the missing cusp problem is indeed a problem of the missing brighter giants ($mk < 15.5$) near SgrA*. Even the issue of the missing brighter giants near the SMBH is less severe compared to previous estimates. Our updated star count in the innermost radial bin will probably decrease the previous estimated number of missing $\sim 100$ giants ($mk < 16$) within the central 0.2 pc (Gallego-Cano et al. 2018). This implies that we do not necessarily need to invoke scenarios like a recent major merger event (Merritt 2010). Rather, scenarios that are in play on shorter timescales ($\sim 10$ Myr, Davies 2017) and cause a selective destruction of red giants (e.g., Murphy 2011; Amaro-Seoane & Chen 2014) are sufficient to explain the observations.

### 5.1. Stellar Orbits and the Cusp

By combining the continuity equation of the stellar orbits in phase space with the expected Maxwellian velocity field of a relaxed system (Quinlan et al. 1995), one can show that the shallower a cusp is, the larger is the fraction of loosely bound or unbound stars (see Section 3.1.3 of Alexander 2005). Current observations show that $\sim 30\%$ of the giants within the central arcsecond are on closer orbits around the SMBH. The region is mostly populated by unbound giants or giants on orbits with larger orbital periods. From our sample of 21 late-type stars within the central arcsecond, a significant acceleration has been measured for S37,[5] S17, S38, S21, and S24 (Gillessen et al. 2017). The stars S20 and S134, for which we have not yet determined an orbit, have high 3D velocities ($\sim 350 \pm 34$ km s$^{-1}$) and plausibly are bound to the SMBH. On the other hand, the brightest giants like S10 and S30 with low velocities are on orbits with longer orbital periods ($\geqslant 200$ yr). The current orbital measurements agree with a shallow cusp hypothesis.

### 6. Summary and Discussion

By combining over a decade of spectroscopic observations, we construct deep spectra of late-type stars within the central arcsecond. We expand the spectral-type-classification in the vicinity of SgrA* down to 17.5 mag, and identify new late-type stars, including the first five warm giants (G2-G8III) in this region. Assuming a metallicity of $Z = 1.5 Z_\odot$, we measure an initial-mass range of 0.5–2 $M_\odot$ and radii of 4–30 $R_\odot$ for these stars. The estimated tidal disruption radii of these stars range within $0\rlap{.}''0004$–$0\rlap{.}''003$. The apparent spread of the late-type stars on the H–R diagram suggests a metallicity spread among these stars.

Our updated late-type star count based on individual spectral classification within the central arcsecond, combined with the reanalyzed data from Bartko et al. (2010), is used to reconstruct the surface density profile of giant stars. We find a cusp structure for the old giant stars ($m_K < 17$) within 0.25–10 arcsec (0.02–0.4 pc). This cusp is described by a single power law with an exponent $\Gamma = 0.34 \pm 0.04$. Our study provides the first spectroscopic detection of a cusp of giant stars toward the central arcsecond. This finding agrees with the recent studies found a cusp profile of fainter giants outside the central arcsecond out to 2 pc from the SgrA$^*$ (Baumgardt et al. 2018; Gallego-Cano et al. 2018; Schödel et al. 2018).

---

[5] By detecting an acceleration in the radial velocity of the star S37 we are able to, for the first time, determine an orbit for this star. The S37 orbital elements are as follows: $a['']= 0.98 \pm 0.05$, $e = 0.45 \pm 0.01$, $i[°] = 111.15 \pm 0.47$, $\Omega\,[°] = 347.26 \pm 1.09$, $t_P[\mathrm{yr}-2000] = 1742.22 \pm 28.87$, and $T[\mathrm{yr}] = 354.77 \pm 27.03$.

The current orbital measurements of late-type S-stars within the central arcsecond agree with a shallow cusp theory. A continuous deep monitoring of fainter late-type stars is the crucial path for confirming the found cusp in this study, and to characterize the kinematics of its habitants.

We thank the anonymous referee for helpful suggestions that significantly improved this Letter. This research has made use of NASA's Astrophysics Data System Bibliographic Services.